# AMR and magnetometry studies of ultra thin GaMnAs films.

**A.W. Rushforth[1], A.D. Giddings, K.W. Edmonds, R.P. Campion, C.T. Foxon, and B.L. Gallagher.**

School of Physics and Astronomy, University of Nottingham, University Park, Nottingham, United Kingdom, NG7 2RD.

We have measured the Anisotropic Magnetoresistance (AMR) of ultra thin (5nm) $Ga_{0.95}Mn_{0.05}As$ films. We find that the sign of the AMR can be positive or negative, which may depend on the direction of the current with respect to the crystal. At low temperatures, transport measurements and SQUID magnetometry suggest that the magnetisation has a component pointing out of the plane of the film.

**1 Introduction** The magnetic and transport properties of ferromagnetic semiconductors are of great interest for their potential applications in magnetic sensing and storage as well as for their utilisation in spintronic devices. In particular, in the dilute magnetic semiconductor GaMnAs, ferromagnetic ordering of the substitutional Mn ions is mediated through a p-d exchange interaction with the itinerant holes. Therefore, the magnetic and transport properties depend upon the carrier density[1,2]. An understanding of these properties is important for determining the underlying physical nature of the magnetism in this system and for future applications.

Recent magnetometry measurements on GaMnAs films (thickness > 400nm) have shown that the magnetic easy axis tends to lie in the plane for compressive strained films and points out of plane for tensile strained films [3]. The difference in the magnetic anisotropy arises from the effect of strain on the splitting of the heavy and light hole bands. For very low hole concentration the compressive strained films also show an out of plane magnetic easy axis when only the lowest, heavy hole subband is occupied.

Electrical transport measurements have provided an alternative method to study the magnetic properties of GaMnAs films [4]. Such measurements exploit the phenomenon of Anisotropic Magnetoresistance (AMR), which arises due to the spin-orbit interaction and leads to a resistance that is dependent upon the angle between the magnetisation and the direction of the current. One advantage of this method is that it allows one to study the magnetic properties of samples with small magnetic moments, which would be difficult to study by conventional magnetometry techniques.

Here we employ electrical transport measurements of the AMR, and also SQUID (superconducting quantum interference device) magnetometry to study the transport and magnetic properties of ultra thin (5nm) $Ga_{0.95}Mn_{0.05}As$ films. We find that the sign of the AMR can be positive or negative, which may depend on the direction of the current with respect to the crystal. Also, the results suggest that the magnetic easy axis has a component pointing out of the plane of the film.

---

[1] Corresponding author email: Andrew.Rushforth@Nottingham.ac.uk

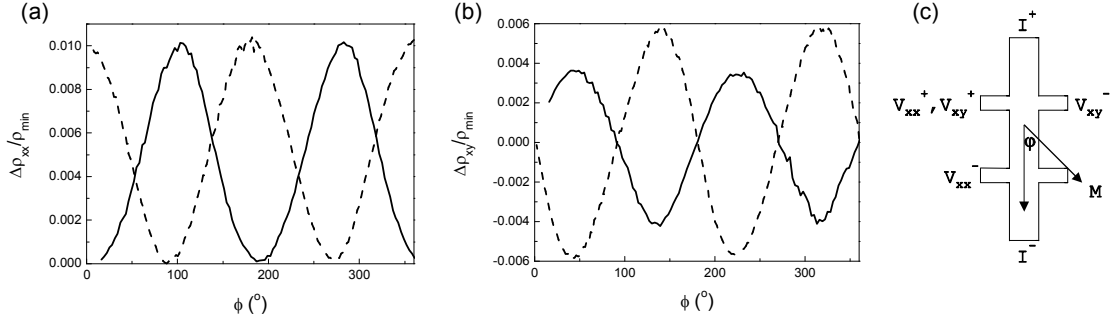

**Figure 1.** (a) $\Delta\rho_{xx}/\rho_{min}$ and (b) $\Delta\rho_{xy}/\rho_{min}$ for 5nm Ga$_{0.95}$Mn$_{0.05}$As films with current applied along the $[100]$ (solid lines) and $[1\bar{1}0]$ (dashed lines) direction. $\phi$ is the angle between a magnetic field of 1T applied in the plane of the film and the direction of the current. T=30K. (c) Configuration of contacts for measurements.

## 2 Experimental details

The 5nm Ga$_{0.95}$Mn$_{0.05}$As film was grown by low temperature (250$^{\circ}$C) molecular beam expitaxy onto a 330nm HT-GaAs buffer layer grown on a semi-insulating GaAs(001) substrate. Standard photolithography techniques were used to fabricate Hall Bars of width 45μm with voltage probes separated by 285μm. Two Hall bars, with the current along the $[100]$ and $[1\bar{1}0]$ directions, were fabricated from positions separated by 5mm on the same wafer. The devices were measured in separate cooldowns and the results for each device were confirmed in a second cooldown. The longitudinal resistance $R_{xx}$ and the Hall resistance $R_{xy}$ were measured using four probe techniques (the configuration is illustrated in Fig. 1(c)) with a DC current of 1μA. Magnetometry measurements were carried out using a commercial Quantum Design SQUID magnetometer.

## 3 Results

Figures 1(a) and (b) show the fractional change in the resistivity, $\Delta\rho_{xx}/\rho_{min}$ and $\Delta\rho_{xy}/\rho_{min}$ ($\Delta\rho_{xx} = \rho_{xx}$-$\rho_{min}$ and $\Delta\rho_{xy} = \rho_{xy}$-$\rho_{min}$) for the two samples when a magnetic field of 1T is applied in the plane of the film and rotated through 360$^{\circ}$. A field of 1T is sufficient to saturate the magnetisation along the direction of the external field at this temperature. Here $\rho_{min}$ is the minimum value of the longitudinal resistivity taken over 360$^{\circ}$. The sample temperature is 30K which is below the Curie temperature, $T_c$ = 52K for this material (see Fig. 3). The simplest form of the AMR for an isotropic material is given by

$$\rho_{xx} = \tfrac{1}{2}(\rho_{||}+\rho_{\perp}) + \tfrac{1}{2}(\rho_{||}-\rho_{\perp})cos2\phi \qquad (1)$$

$$\rho_{xy} = \tfrac{1}{2}(\rho_{||}-\rho_{\perp})sin2\phi \qquad (2)$$

where $\phi$ is the angle between the magnetisation and the current direction and $\rho_{||}$ and $\rho_{\perp}$ are the resistivities when the magnetisation is parallel or perpendicular to the current direction respectively. For most ferromagnetic metals $\rho_{||} > \rho_{\perp}$ [5]. However, for compressive strained GaMnAs $\rho_{||} < \rho_{\perp}$ is usually observed and the relations in equations (1) and (2) have been found to hold well[6,7]. Figure 1 shows that, for these 5 nm films the sign of ($\rho_{||}$-$\rho_{\perp}$) can be positive or negative. Furthermore, the sign of the oscillation in $\Delta\rho_{xy}/\rho_{min}$ is opposite to that for $\Delta\rho_{xx}/\rho_{min}$ and they differ in magnitude also. This shows that, for these 5nm Ga$_{0.95}$Mn$_{0.05}$As films the relations in equations (1) and (2) do not hold. This limited data set indicates that the AMR may have a strong dependence on the orientation of the current with respect to the crystal. Further evidence for this has been provided by a recent observation of ($\rho_{||}$-$\rho_{\perp}$)>0 in a different 5nm Ga$_{0.95}$Mn$_{0.05}$As film with the current along the [110] direction, although a discussion of this observation was not presented[8]. Previous studies of ferromagnetic metals have shown that, for crystalline material, the AMR can depend on the angle of M and I with respect to the crystal[5]. Recently, the magnetocrystalline contribution to the AMR was used to explain a 10-15% difference in the amplitudes of $\rho_{xy}$ and $\rho_{xx}$ for a 50nm $Ga_{0.97}Mn_{0.03}As$ film[4]. The results presented here may indicate that the magnetocrystalline contribution dominates the AMR in these 5nm $Ga_{0.95}Mn_{0.05}As$ films.

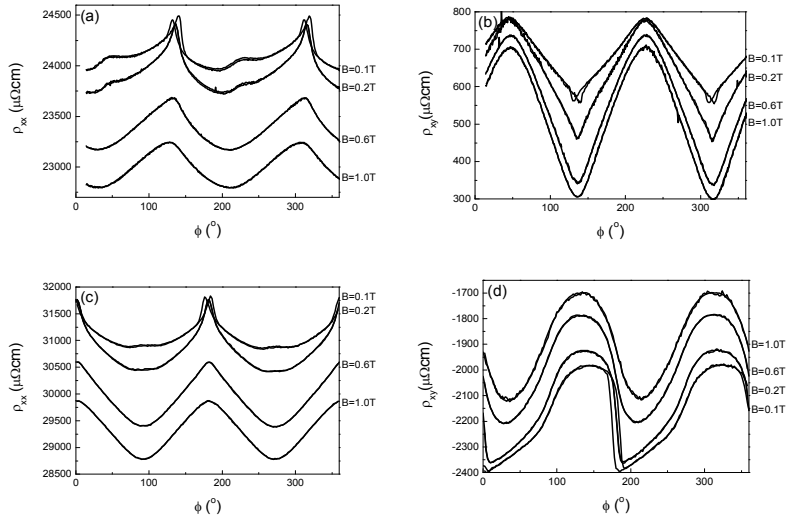

**Figure 2** $\rho_{xx}$ and $\rho_{xy}$ for 5nm $Ga_{0.95}Mn_{0.05}As$ films with current applied along the $[100]$ ((a) and (b)) and $[1\bar{1}0]$ ((c) and (d)) direction. T=4.2K

At lower temperatures we observe further deviations from the usual form for AMR. Figure 2 shows $\rho_{xx}$ and $\rho_{xy}$ measured at 4.2K for the two samples with magnetic fields of 1T, 0.6T, 0.2T and 0.1T applied in the plane of the film and rotated through 360°. There are several points to observe in this data. Firstly, the increase in the average of $\rho_{xx}$ as the field is reduced is a manifestation of the isotropic magnetoresistance, which has been observed in many $Ga_{1-x}Mn_xAs$ samples, and has been attributed to weak localisation and spin-disorder scattering at low and high temperatures respectively [9,10]. Secondly, the fact that the average of $\rho_{xy}$ is offset from zero is due to some component of the longitudinal voltage appearing in the transverse measurement. This can arise due to sample inhomogeniety. Thirdly, the most striking observation is the distortion of the shape of the AMR curves which becomes more pronounced as the field is reduced. Both the $\rho_{xx}$ and $\rho_{xy}$ curves develop hysteresis for fields below 0.2T indicating that the magnetisation is not following the external magnetic field precisely. The distortions of the curves from the usual form for the AMR tend to occur at angles where the magnetisation crosses the $[110]$ and $[1\bar{1}0]$ directions. At low temperatures the magnetic easy axis tends to show biaxial symmetry lying in the plane along the $[100]$ and $[010]$ directions for compressive strained $Ga_{1-x}Mn_xAs$ films[11], making the $[110]$ and $[1\bar{1}0]$ directions the in-plane hard magnetic directions. This would suggest that at low temperatures and low magnetic fields the magnetisation may not track the external field and may even have a component pointing out of the plane of the film when the external field crosses the $[110]$ and $[1\bar{1}0]$ directions.

We have also perfomed SQUID magnetometry measurements on the 5nm $Ga_{0.95}Mn_{0.05}As$ film. Figure 3 shows the remnant magnetisation measured along several crystallographic directions as a function of temperature after cooling down in a field of 1000Oe. For the $[001]$ direction, which is the direction perpendicular to the plane of the film, there is a small, but significant remnance. The magnitude of this remnance at 2K indicates that the magnetisation could be at an angle of 10° to the film plane. This would be larger than any experimental error in mounting the sample. Figure 3 also shows that the magnetic easy axis is along the $[1\bar{1}0]$ direction over the whole temperature range for this sample. This is in contrast to the transport measurements in Fig.2 which suggest that the $[110]$ and $[1\bar{1}0]$ directions are both hard magnetic directions at 4.2K. This discrepancy may be due to the fact that the samples for transport and magnetometry measurements were taken from slightly different positions on the wafer and so may have slightly different magnetic properties due to inhomogeniety across the wafer. Indeed, $Ga_{1-x}Mn_xAs$ films have been observed to show a crossover in plane biaxial to in plane uniaxial anisotropy as the temperature is increased[11].

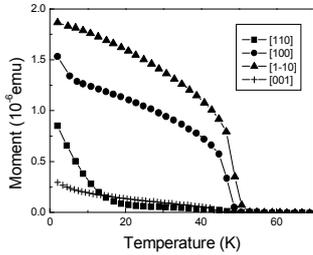

**Figure 3** Remnant moment measured by SQUID magnetometer. The sample is first cooled from room temperature in a field of 1000 Oe. The field is then removed and the moment is measured as a function of temperature.

We have measured the hole density of our samples by measuring $\rho_{xx}$ and $\rho_{xy}$ in magnetic fields up to 12T and carrying out analysis previously used for determining carrier densities in $Ga_{1-x}Mn_xAs$ films[12]. We find $p=1.6x10^{20}cm^{-3}$, which is approximately half the value measured for thicker films with the same nominal Mn content. This may indicate that the growth of these ultra thin films results in the incorporation of a lower number of Mn ions on substitutional sites. This might arise due to temperature instability during growth which would normally stabilise during the growth of thicker films. Alternatively, the apparent low density could be accounted for if the electrical thickness were less than 5nm due to band bending near the surface and GaAs interface.

In summary, transport and magnetometry measurements on 5nm $Ga_{0.95}Mn_{0.05}As$ films reveal several properties which are different to those usually observed in thicker (>25nm) $Ga_{0.95}Mn_{0.05}As$ films. These include the AMR having a positive or negative sign which may depend upon the direction of the current with respect to the crystal. There is also evidence that the magnetisation has a component pointing out of the plane of the film. At this stage it is not clear whether these properties arise due to the thickness of the films or due to a reduced Mn content and carrier density.

**Acknowledgements**   We would like to thank M. Sawicki, T. Jungwirth and K. Výborný for useful discussions. We are grateful for financial support from the EPSRC (GR/S81407/01).